\documentclass{article}
\usepackage{amssymb}
\usepackage{amsfonts}
\usepackage{amsmath}
\usepackage{graphicx}% Include figure files
\begin{document}

\title{Spectral Analysis on Explosive Percolation}

\author{N. N. Chung$^1$, L. Y. Chew$^2$ and C. H. Lai$^{3,4,5}$ \\
\\
$^{1}$ Temasek Laboratories,\\
 National University of Singapore, \\
 Singapore 117508 \\
$^{2}$ School of Physical \& Mathematical Sciences, \\
Nanyang Technological University,\\
 21 Nanyang Links, Singapore 637371 \\
$^{3}$Beijing-Hong Kong-Singapore Joint Centre for \\
Nonlinear and Complex Systems (Singapore), \\
National University of Singapore, \\
Kent Ridge 119260, Singapore \\
$^{4}$ Department of Physics, \\
National University of Singapore,\\
 Singapore 117542 \\
 $^5$Yale-NUS College, \\
 6 College Avenue East, \\
 Singapore 138614}
\maketitle

\begin{abstract}
We study the spectral properties of the process of explosive percolation. In particular, we explore how the maximum eigenvalue of the adjacency matrix of a network which governs the spreading efficiency evolves as the density of connection increases. Interestingly, for networks with connectivity that grow in an explosive way, information spreading and mass transport are found to be carried out inefficiently. In the conventional explosive percolation models that we studied, the sudden emergences of large-scale connectivity are found to come with relatively lowered efficiency of spreading. Nevertheless, the spreading efficiency of the explosive model can be increased by introducing heterogeneous structures into the networks.
\end{abstract}

\newpage

In the past few years, there has been intensive research on explosive percolation - the abrupt development of large-scale connectivity in networks, immediately after Achlioptas et. al. brought out their percolation model with an unexpected sharp transition \cite{Achlioptas09}. In the model, when one increases progressively the number of connections between nodes in a network following the suppression principle \cite{Cho11}, above some critical threshold, a giant connected cluster emerges suddenly. The resulting transition was originally thought to be discontinuous.   It is shown however to be continuous for some models later \cite{Costa10,Grassberger11,Andrade11,Lee11,Riordan11}. In addition to the explosive percolation of a single giant connected component, the abrupt creation of stable multiple giant components are shown to be possible as well \cite{Chen11,Bohman04}. More recently, strongly and weakly discontinuous transitions are introduced to classify how explosive a percolation transition is \cite{Nagler11}. By now, the study of discontinuous percolation transition has been extended to scale-free networks \cite{Cho09,Radicchi09}, real-world networks with community structure \cite{Pan11}, two- and higher-dimensional lattices \cite{Chae12,Schrenk11,Ziff10,Araujo10}, etc. Concurrently, other approaches like the weighted rule \cite{Cho10} and the Hamiltonian approach \cite{Moreira10} have also been proposed to obtain similar transitions. Most notably, this process has shown to be a possible mechanism that explains the growth process of a couple of real networks, including the human protein homology network \cite{Rozenfeld10} and nanotube clustering \cite{Kim10}. For real-world systems, such behavior can have vital consequences when the addition of single links \cite{Cho10,Manna11} may drastically change macroscopic connectivity and hence the dynamic and function of the networks.  For instance, in a neuronal circuit, this would mean that the growth of one or a few additional synaptic connections might drastically alter the information processing function in the brain. Similarly, the establishment of a small number of specific social relations may significantly increases the possible extent of infectious diseases or rumors. Here, we study the evolution of spectral properties of the network during the process of explosive percolation. In particular, we explore how the maximum eigenvalue of the adjacency matrix of the network evolves as the density of connection increases. This is of interest as the maximum eigenvalue of the network governs how information or diseases spread \cite{Boguna02,Kim07}. If it increases explosively due to the sudden emergence of a giant component, the spreading process will be enhanced. On the other hand, if the maximum eigenvalue remains small at the percolation threshold, an efficient spreading is not possible despite the existence of a giant connected cluster in the network.

The smallest cluster (SC) model is one of the simplest model that show strongly discontinuous percolation transition. It starts with $N=2^n$ isolated nodes. At each step, the two smallest clusters in the network are identified and merged into a larger cluster through the creation of a link between them. In this case, $N/2$ clusters of size $2$ are created during the first phase, i.e. the first $N/2$ steps. In the second phase, $N/4$ links are added, each connecting two clusters of size $2$ into a cluster of size $4$. The process continues and by the end of the phase $y=n-1$, only two components remain, each with size $N/2$. Then, the addition of the next edge connects these two components, resulting in a jump in value of $N/2$  in the size of the largest component ($S$). If the nodes to be connected in each step are chosen randomly from the two smallest clusters, then at the end of phase $y$, since $N/2^y$ edges are added to the network and the total degree is increased by $N/2^{(y-1)}$, the evolution of the degree distribution will be described by:
\begin{equation}
\partial P(k,y)/ \partial y = \frac{1}{2^{y-1}} (-P(k,y) + P(k-1,y)) \,. \label{partialP}
\end{equation}
With this, we can obtain the probability of nodes with degree $y$ at phase $y$:
\begin{equation}
P(y,y)=\frac{1}{2^{\sum_{j=0}^{y-1} j}} \,.
\end{equation}
Note that $P(k,y)=0$ for all $k>y$.
In the earlier phases, the maximum degree of the network, $k_m$ increases linearly with the phase. Nonetheless, since $P(y,y)$ drops rapidly as $y$ increases, beyond a critical phase, the probability for $k_m$ to increase becomes very small. Here, we define $y_c$ to be the critical phase where $P(y_c,y_c)=1/N$. The value of $y_c$ can then be obtain by solving the following equation:
\begin{equation}
y_c (y_c-1) = 2n \,.
\end{equation}
Given that the maximum eigenvalue ($\lambda_m$) of a network's adjacency matrix is inversely proportional to the square root of the network's maximum degree, $\lambda_m$ is not expected to grow explosively at the percolation threshold as the size of giant component does. In fact, as shown in Fig. 1(a) for a smaller network size and Fig. 1(b) for a larger network size, the maximum eigenvalues do not increase further after the critical phases. Note that the results are obtained through numerical iteration of Eq. (\ref{partialP}) with the maximum eigenvalue being approximated as the maximum eigenvalue of the largest cluster \cite{Chung12}:
\begin{equation}
\lambda_m =\sqrt{k_m + \frac{\langle k ^2\rangle}{\langle k \rangle ^2}\left(\langle k ^2\rangle - \frac{k_m^2}{S} \right) } \,.
\end{equation}
For the smaller network size, a result based on an averaging of $20$ network realizations is also obtained for comparison.

On the other hand, if the nodes to be connected in each step are chosen to be the largest-degree nodes in the two smallest clusters, then at the end of phase $y$,
\begin{equation}
\partial P(k,y)/ \partial y = \begin{cases} - \frac{1}{2^{y-1}} , & \mbox{for} \, k=k_m \,, \\  \frac{1}{2^{y-1}}, & \mbox{for} \, k=k_m+1 \,, \\ 0, & \mbox{for all other} \, k \,. \end{cases}   \label{partialP2}
\end{equation}
In this modified SC model, the maximum degree increases linearly with the phase. Thus, as shown in Fig. 1, larger maximum eigenvalues and more efficient spreading are obtainable at the percolation threshold for giant clusters that emerge explosively.

Next, we look at a more general model, the Gaussian model \cite{Araujo10}. In this model, the distribution of the cluster size is controlled to be Gaussian by implementing the following probability of occupation:
\begin{equation}
p = \exp \left[ -\alpha \left(\frac{s-\bar{s}}{\bar{s}}\right)^2\right]  \label{GaussianP}
\end{equation}
for an edge that is selected randomly among the empty ones. In the equation, $s$ denotes the size of the cluster that would be formed by adding the selected edge while $\bar{s}$ denotes the average cluster size after adding the edge. When an intra-cluster link is chosen, $s$ will be taken as twice the cluster size. The parameter $\alpha$ controls the size dispersion and we consider only for cases where $\alpha \geq 0$. For $\alpha=0$, all edges have the same probability to be added and the model gives the Erd\H{o}s-R\'{e}nyi graph. In this case, a cluster of relative size is expected to appear at the percolation threshold $t_c =0.5$ and grows gradually as the connection density increases (see the inset of Fig. 2). For $\alpha > 0$, the formation of a cluster with size which differs significantly from the average cluster size is suppressed. The result is a sudden emergence of a giant connected cluster at $t_c \approx 1$. In the inset of Fig. 2, we show the explosive percolation transitions for $\alpha=0.5$ and $\alpha=2$.

For this model with $\alpha>0$, we consider only the addition of inter-cluster connections for the derivation of the evolution of the network distribution. Letting $\tau$ to be the ratio of the total number of sampled edges to the network size, then at $\tau$, the sampling probability for an edge between a node in the cluster of size $s$ and a node in the cluster of size $s'$ is $2 Q(s,\tau) Q(s',\tau)$ where $Q(s,\tau)  = \sum_k P(s,k,\tau)$. The probability for this edge to be added is $\exp \left[-\alpha \left( \frac{s+s'-\bar{s}}{\bar{s}} \right) \right] $. If this edge is added, $P(s,k,\tau)$ will decrease by $s P(s,k,\tau)/N Q(s,\tau)$. On the other hand, to form a cluster of size $s$, the selected nodes must be in clusters of sizes $u$ and $v$ with $u+v=s$. The probability for a connection to be added between these two nodes is $\exp \left[-\alpha \left( \frac{s-\bar{s}}{\bar{s}} \right) \right]$. If this edge is added, nodes with degree $k$ in the two clusters will increase the fraction of nodes with cluster size $s$ and degree $k$ unless they are the selected nodes. Then again, nodes with degree $k-1$ in the two clusters will increase the fraction of nodes with cluster size $s$ and degree $k$ if they are the selected nodes. The evolution of the distribution of the network can thus be written as:
\begin{eqnarray}
\nonumber
\frac{\partial P(s,k,\tau)}{\partial \tau} &=& -2 \frac{s}{N} P(s,k,\tau) \sum_{s'} Q(s',\tau) \, e^{\left[-\alpha \left( \frac{s+s'-\bar{s}}{\bar{s}} \right) \right] }\\
\nonumber
 & +&  \sum_{u+v=s} Q(u,\tau) Q(v,\tau) \, e^{\left[-\alpha \left( \frac{s-\bar{s}}{\bar{s}} \right) \right]} \\
 \nonumber & & \left( \frac{(u-1)P(u,k,\tau)}{N Q(u,\tau)}  + \frac{P(u,k-1,\tau)}{N Q(u,\tau)}  \right.\\
 & & \left. + \frac{(v-1)P(v,k,\tau)}{N Q(v,\tau)}  + \frac{P(v,k-1,\tau)}{N Q(v,\tau)} \right) \,. \label{partialPsk}
\end{eqnarray}
Note that the ratio of the total number of added edges and the network size can be approximated as $ t= \langle k \rangle/ 2$.

Equation (\ref{partialPsk}) is iterated numerically, then the degree distribution $R(k,\tau)  = \sum_s P(s,k,\tau)$ is used to calculate $\lambda_m$ for $\tau$ with $ \sum_k R(k,\tau)=1$. The growth of $\lambda_m$ versus $t$ is shown in Fig. 2 for both the random network and the explosive model. As shown, $\lambda_m$ increases gradually as $t$ increases. Interestingly, while the size of the largest connected component at $t \approx 1$ is considerably larger for the explosive models, the maximum eigenvalue is found to be consistently smaller throughout the process. In other words, although the connected component is larger in networks generated through the explosive model, the efficiency of spreading in these networks is lower compared to that of the random graphs. In addition, the sudden emergence of a large-scale connectivity in the explosive model comes at a cost of lowered navigation efficiency in the network. In Fig. 3, the growth of average inverse path length ($L$) are shown for networks developed through the Gaussian model with $\alpha=0$, $0.5$ and $2$. For $\alpha>0$, although the average inverse path length increases explosively at the percolation threshold, it is still significantly smaller than that of the random networks.

In order to increase the spreading efficiency of the networks, we implemented a modified Gaussian model. At each time step, instead of drawing two nodes randomly from the network, nodes $i$ and $j$ are drawn from the network with probabilities $w_i / \sum_m w_m$ and $w_j / \sum_m w_m$ respectively. Here, $w_m$ are weights assigned to nodes in the network following the Chung and Lu (CL) model \cite{Cho09,Chung02} which is used to construct artificial scale-free networks. Specifically, the nodes are indexed as $i=1, 2, \cdots, N$ and each node is assigned a weight of
\begin{equation}
w_i = (i+i_0-1)^{-\mu}
\end{equation}
where
\begin{equation}
i_0 = \begin{cases} 1 , & \mbox{for} \, \mu<0.5 \,, \\  N^{1-\frac{1}{2 \mu}}, & \mbox{for} \, 0.5<\mu<1  \,. \end{cases}
\end{equation}
Then, if edge $ij$ does not already exist, it will be added with the probability described by Eq. (\ref{GaussianP}).
Numerical simulations are performed for this modified model. Indeed, as shown in Fig. 4, $\lambda_m$ is greatly increased when heterogeneous structures are introduced into the explosive model.

Explosive percolation model has altered the traditional understanding on percolation transitions. While the sudden emergence of a large-scale connectivity may significantly change the dynamic and function of a network, for information and mass transport in real-world networks, the existence of a large-scale connectivity alone is not sufficient. The fact that most of the real-world networks are small-world networks, in which the average path length is scaled as log($N$), reveals that navigation efficiency in a network is a critical consideration. However, as shown by our results, the sudden emergence of a giant component in the explosive model comes at a cost of lowered spreading efficiency in the network. As such, the insertion of heterogeneous structures into the network becomes important in achieving high spreading efficiency in explosive models. \\
\\

\noindent
{\bf \large{Acknowledgement} } \\

This work is supported by the Defense Science and Technology Agency of Singapore under project agreement of POD0613356.

\newpage

\begin{figure}
\begin{center}
\includegraphics[scale=0.6]{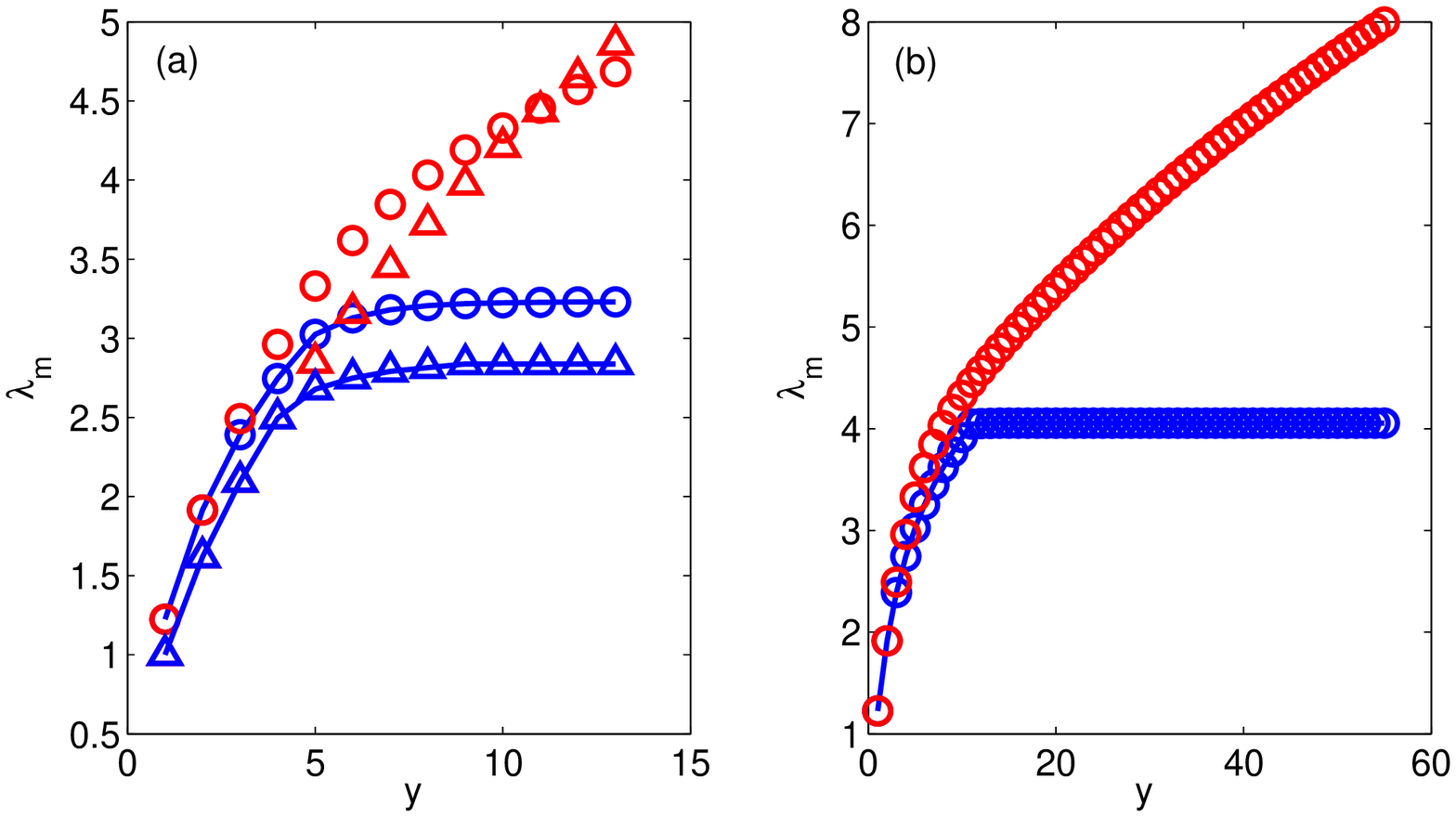}
\end{center}
\caption{The maximum eigenvalue at the end of each phase for the smallest cluster model (lower curves) and the modified smallest cluster model (upper curves) of size (a) $N=2^{13}$ and (b) $N=2^{55}$. Note that the ensemble averages obtained by simulation of the SC model with $2^{13}$ nodes are shown as triangles whereas the approximations obtained through numerical iteration of Eq. (\ref{partialP}) and (\ref{partialP2}) are shown as circles.} \label{fig1}
\end{figure}

\begin{figure}
\begin{center}
\includegraphics[scale=0.5]{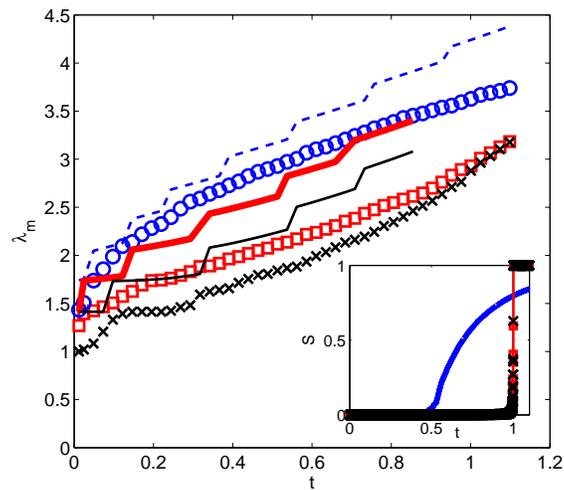}
\end{center}
\caption{The maximum eigenvalue versus the ratio of the added edges for the Gaussian model with $\alpha=0$, $0.5$ and $2$. Note that the numerical results are obtained through an average of $20$ network realizations of size $N=2^{13}$ and are shown as circles ($\alpha=0$), squares ($\alpha=0.5$) and crosses ($\alpha=2$) whereas the analytical approximations are shown as dashed ($\alpha=0$), thick ($\alpha=0.5$), and solid ($\alpha=2$) lines. For numerical iteration of Eq. (\ref{partialPsk}), the parameter used are $s=1,2,\cdots,N$, $k=0,1,\cdots,25$ and $0<t<0.85$. Insets: The percolation transition for the Gaussian model with $\alpha=0$ (solid line), $0.5$ (line with squares) and $2$ (crosses).} \label{fig2}
\end{figure}

\begin{figure}
\begin{center}
\includegraphics[scale=0.5]{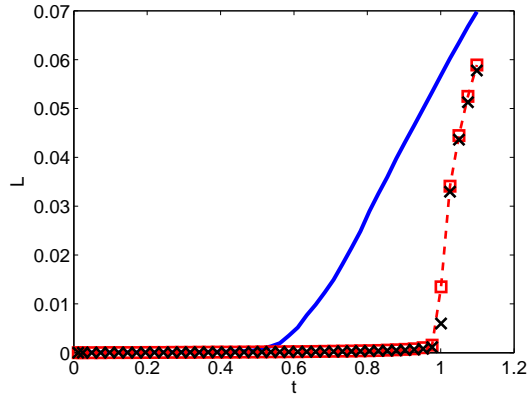}
\end{center}
\caption{The average inverse path length versus the connection density for the Gaussian model with $\alpha = 0$ (thick line), $0.5$ (squares with dashed line) and $2$ (crosses). Note that results are obtained by averaging over $20$ network realizations with $N=2^{13}$.} \label{fig3}
\end{figure}

\begin{figure}
\begin{center}
\includegraphics[scale=0.5]{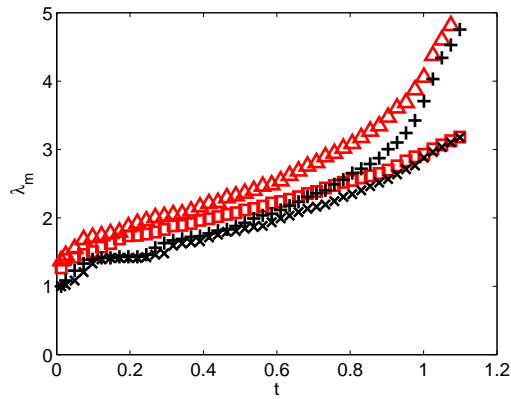}
\end{center}
\caption{The maximum eigenvalue versus the ratio of the added edges for the Gaussian and modified Gaussian model. Note that the numerical results are obtained through an average of $20$ network realizations of size $N=2^{13}$ and are shown as squares (Gaussian model with $\alpha=0.5$), triangles (modified Gaussian model with $\alpha=0.5$), crosses (Gaussian model with $\alpha=2$) and plus signs (modified Gaussian model with $\alpha=2$).} \label{fig4}
\end{figure}

\end{document}